\documentclass[fleqn,twoside,twocolumn,nofootinbib]{revtex4} 
\usepackage{ujp} 
\begin{document}
\title[GLUCOSE OXIDASE IMMOBILIZATION ONTO CARBON NANOTUBE NETWORKING]
{GLUCOSE OXIDASE IMMOBILIZATION ONTO CARBON NANOTUBE NETWORKING}%
\author{V.A. KARACHEVTSEV}
\affiliation{B. Verkin Institute for Low Temperature Physics and
Engineering,\\ Nat. Acad. of Sci. of Ukraine}
\address{47, Lenin Ave., Kharkov 61103, Ukraine}
\author{A.Yu. GLAMAZDA}%
\affiliation{B. Verkin Institute for Low Temperature Physics and
Engineering,\\ Nat. Acad. of Sci. of Ukraine}%
\address{47, Lenin Ave., Kharkov 61103, Ukraine}%
\author{E.S.~ZARUDNEV}
\affiliation{B. Verkin Institute for Low Temperature Physics and
Engineering,\\ Nat. Acad. of Sci. of Ukraine}%
\address{47, Lenin Ave., Kharkov 61103, Ukraine}%
\author{M.V.~KARACHEVTSEV}%
\affiliation{B. Verkin Institute for Low Temperature Physics and
Engineering,\\ Nat. Acad. of Sci. of Ukraine}%
\address{47, Lenin Ave., Kharkov 61103, Ukraine}%
\author{V.S.~LEONTIEV}
\affiliation{B. Verkin Institute for Low Temperature Physics and
Engineering,\\ Nat. Acad. of Sci. of Ukraine}
\address{47, Lenin Ave., Kharkov 61103, Ukraine}
\author{A.S.~LINNIK}%
\affiliation{B. Verkin Institute for Low Temperature Physics and
Engineering,\\ Nat. Acad. of Sci. of Ukraine}
\address{47, Lenin Ave., Kharkov 61103, Ukraine}
\author{O.S.~LYTVYN}
\affiliation{V. Lashkaryov Institute of Semiconductor Physics, Nat.
Acad. of Sci. of Ukraine}%
\address{41 Nayki Ave., 03028 Kyiv, Ukraine}%
\author{A.M.~PLOKHOTNICHENKO}%
\affiliation{B. Verkin Institute for Low Temperature Physics and
Engineering,\\ Nat. Acad. of Sci. of Ukraine}
\address{47, Lenin Ave., Kharkov 61103, Ukraine}
\author{S.G.~STEPANIAN}%
\affiliation{B. Verkin Institute for Low Temperature Physics and
Engineering,\\ Nat. Acad. of Sci. of Ukraine}%
\address{47, Lenin Ave., Kharkov 61103, Ukraine}%
\udk{???}  \razd{\secix}

\setcounter{page}{700}%
\maketitle

\begin{abstract}
When elaborating the biosensor based on single-walled carbon
nanotubes (SWNTs), it is necessary to solve such an important
problem as the immobilization of a target biomolecule on the
nanotube surface. In this work, the enzyme (glucose oxidase (GOX))
was immobilized on the surface of a nanotube network, which was
created by the deposition of nanotubes from their solution in
1,2-dichlorobenzene by the spray method. 1-Pyrenebutanoic acid
succinimide ester (PSE) was used to form the molecular interface,
the bifunctional molecule of which provides the covalent binding
with the enzyme shell, and its other part (pyrene) is adsorbed onto
the nanotube surface. First, the usage of such a molecular interface
leaves out the direct adsorption of the enzyme (in this case, its
activity decreases) onto the nanotube surface, and, second, it
ensures the enzyme localization near the nanotube. The comparison of
the resonance Raman (RR) spectrum of pristine nanotubes with their
spectrum in the PSE environment evidences the creation of a
nanohybrid formed by an SWNT with a PSE molecule which provides the
further enzyme immobilization. As the RR spectrum of an SWNT:PSE:GOX
film does not essentially differ from that of SWNT:PSE ones, this
indicates that the molecular interface (PSE) isolates the enzyme
from nanotubes strongly enough. The efficient immobilization of GOX
along the carbon nanotubes due to PSE is confirmed with atom-force
microscopy images. The method of molecular dynamics allowed us to
establish the structures of SWNT:PSE:GOX created in the aqueous
environment and to determine the interaction energy between hybrid
components. In addition, the conductivity of the SWNT network with
adsorbed PSE and GOX molecules is studied. The adsorption of PSE
molecules onto the SWNT network causes a decrease of the
conductivity, which can be explained by the appearance of scattering
centers for charge carriers on the nanotube surface, which are
created by PSE molecules.
\end{abstract}

\section{Introduction}

Due to their unusual physical, optical, thermal, and electronic
properties, single-walled carbon nanotubes (SWNTs) have a huge
potential of different promising applications including the
biosensing. The main challenge in the development of the devices is
the biofunctionalization of nanomaterial surfaces and the creation
of appropriate interfaces between the nanotubes and the biosystems.
Carbon-nanotube-based biological sensors could find applications,
for example, in measuring the concentrations of glucose in blood.
This is particularly important because the number of diabetics in
the world increases continuously and dramatically.

In the last few years, the applications of conducting properties of
carbon nanotubes in the biological sensoring have demonstrated the
high efficiency of this approach, despite the elaborations being in
their infancy [1--3]. The use of the conductivity of carbon
nanotubes as the detection and measuring method in midget biosensors
showed some preferences over the optical methods currently being
employed  in the clinical work. This is related to the fact that the
most of biological processes involve electrostatic interactions
and/or a charge transfer, which can be directly detected with the
electronic equipment. Carbon nanotubes can be easily integrated into
the electronic device. Moreover, the average size of an nanotube is
usually compatible with the molecular size of a compound being
analyzed, which increases the sensitivity of the measurement. The
conductivities of individual nanotubes or carbon nanotube networks
can be utilized in the investigation and the development of
biosensors [4, 5]. Networks formed by hundreds or thousands of
carbon nanotubes distributed randomly between metallic contacts have
a larger active area for the detection and can operate at higher
currents. The technology for the fabrication of carbon-nanotube
networks is well developed and does not require the expensive
equipment to operate. The networks can be produced from solutions by
deposition onto various substrates.\looseness=1

When working out biological sensors, in which an SWNT network is
used, it is necessary to solve such an important problem as the
immobilization of a recognition biomolecule on the nanotube surface.
In researches, an enzyme immobilized on a nanotube is often used as
the recognition element. In some works, the enzyme was immobilized
directly onto the nanotube surface [6]. However, the recent work [7]
showed that the activity of two enzymes (R-chymotrypsin and soybean
peroxidase) decreased significantly after their adsorption onto the
surface of single-walled carbon nanotubes. Thus, the problem of
enzyme immobilization on a nanotube that needs to be solved is to
retain the enzyme native activity despite its immobilization on the
nanotube surface. The problem is closely related to the nanotube
functionalization. Some success on this front has been achieved
owing to the use of a molecular anchor [8, 9] or the
polymer-wrapping of nanotubes [6]. The efficient non-covalent carbon
nanotube functionalization by organic molecules for the
biocompatibility testing was suggested by Chen and co-workers [8].
Their approach utilizes a bifunctional molecule containing
succinimidyl ester and a pyrene moiety to bind proteins to the
nanotube surface. Pyrene attaches to the nanotube surface by means
of the $\pi -\pi $ stacking and the hydrophobic interaction and does
not disrupt the nanotube backbone. Another fragment of the anchor
molecule is succinimidyl ester that binds enzymes to the nanotube
surface. This molecular interlayer provides a sufficiently strong
attachment of the enzyme to the nanotube surface and leaves the
enzyme activity unaffected. The enzyme localization near the
nanotube needs to provide the reliable detection of the charge,
which appears as a result of the biochemical reaction of enzymes
with the probe.

In the present work, the enzyme (glucose oxidase (GOX)) was immobilized onto the
surface of a nanotube network, which was created by the deposition of nanotubes
from their solution in dichlorbenzene by the spray method. 1-pyrenebutanoic
acid N-hydroxysuccinimide ester (PSE) was used to form the molecular interface.
The comparison of the resonance Raman (RR) light scattering spectrum of pristine
nanotubes with their spectrum in the PSE environment evidences the creation of
a nanohybrid formed by SWNT with a PSE molecule which ensures the further enzyme
immobilization. As the RR spectrum of an SWNT:PSE:GOX film does not essentially
differ from that of SWNT:PSE ones, this indicates that the molecular interface (PSE)
isolates the enzyme from nanotubes efficiently. The immobilization of GOX near
a carbon nanotube due to PSE is confirmed by atom-force microscopy
(AFM). The method of molecular dynamics allows us to establish the structures of SWNT:PSE:GOX
and to determine the energies of the intermolecular interaction between
components of the triple complex in the aqueous environment. In addition, we study the
conductivity of the SWNT network with adsorbed PSE and GOX molecules.
The adsorption of PSE molecules on the SWNT network results in a decrease of the
conductivity, which is most likely induced by the appearance of
scattering centers for charge carriers in nanotubes.

\section{Experimental}

SWNTs have been produced by the CoMoCAT method
(SouthWest NanoTechnologies Inc., USA). This method yields nanotubes
with narrow diameters (0.75--0.95 nm) and chirality (6,5)
[10]. PSE was purchased from Sigma-Aldrich, Europe.
All compounds have been used without additional purification.

Nanotube suspensions were prepared in dichlorobenzene
(Sigma-Aldrich, Europe), which is the most efficient organic solvent
for nanotubes. Suspensions were obtained by ultrasonication (44 kHz,
UZDN-2, Sumy, Ukraine) and by the following ultracentrifugation (up
to 18\,000~g). As a result, all insoluble thick bundles of nanotubes,
as well as the metallic catalyst, were precipitated. To get one
sample, we used 0.2 mg of nanotubes, being dissolved in 4 ml of
dichlorobenzene.

The carbon nanotube network on the substrate (quartz) was created by the
spray method. To form this network, a special electronic device was
constructed to control the
nanotube network density, by using the conductivity of the deposited network and a quartz
microbalance. This density was regulated by the variation of the deposition time, as well as
with the nanotube concentration in a solution. After the drying of such a nanotube
network, two gold contact areas with 10 $\mu $m gap between them were
thermodeposited. Glucose oxidase (GOX) was immobilized onto the carbon
nanotube network through PSE, which formed a molecular interface between
nanotubes and the enzyme [8]. PSE was adsorbed from its solution in methanol
(10$^{-4}$ M), and then GOX from its solution in water (10$^{-4}$ M) was
deposited.

The GOX immobilization onto carbon nanotubes was controlled with
an AFM Nanoscope D3000 (Digital
Instruments,USA), as well as with RR spectroscopy. For AFM
measurements, nanotubes were sprayed on mica. For Raman
measurements, three dried films were prepared: pristine nanotubes,
nanotubes with PSE, and SWNT:PSE:GOX. To prepare the film of
nanotubes with adsorbed PSE molecules, SWNTs were mixed with PSE in
methanol (with 1:1 weight ratio, 0.3 mg/ml). The mixture was treated
with the sonication (1~W, 44~kHz) for 30 min. Then the suspension
was deposited on the quartz substrate and dried under a stream of
warm air. SWNT:PSE:GOX hybrids were prepared by the adsorption of GOX
from an aqueous solution on an SWNT:PSE film. Raman experiments have been
performed in the 90$^{\circ}$ scattering configuration relative to
the laser beam, by using 632.8~nm (1.96~eV) light from a He-Ne laser. The
spectra were analyzed using a Raman double monochromator with the
reverse dispersion of 3{\AA}/mm and were detected with a
thermocooled CCD camera. The position of peaks of the bands in the RR
spectrum of a nanotube film was determined with the accuracy not
worse than 0.3 cm$^{-1}$. This level of accuracy has been achieved
due to the observation of the plasma lines of a laser in the spectra
in a vicinity of the bands corresponding to the
tangential mode (G) and the radial breathing mode (RBM) of nanotubes,
which were used in the internal calibration of a spectrometer.

\begin{figure}
\includegraphics[width=7.5cm]{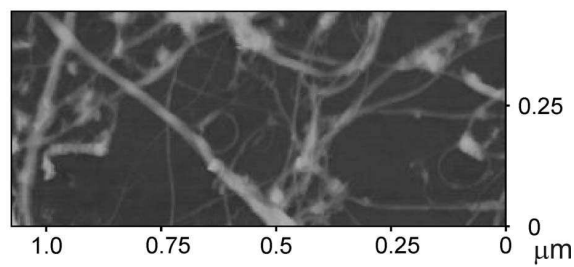}
\vskip-3mm\caption{AFM image of the carbon nanotube network on a
mica substrate obtained by the spray method from an SWNT solution in
dichlorobenzene }
\end{figure}

Volt-ampere characteristics of the nanotube network were determined
by the home-built set-up based on a microcontroller, by means of
which the required range of voltages was formed.
A microamperemeter, which measured a
current through the network, was connected with a computer. A minimal step of the output voltage was 2~mV,
and the current sensitivity reached 0.1~nA.

SWNT hybrids with PSE and with GOX were modeled by the molecular
dynamics method employing NAMD programs [11]. In these
calculations, the force field Charmm27 was used [12]. For the
enzyme, the standard force parameters were applied, and the force
parameters of aromatic carbon were given to carbon nanotube atoms.
Glucose oxidase was obtained from the Protein Data Bank. Its
structure parameters were determined earlier [13]. A PSE molecule
has no standard parameters in this force field. Therefore, an
additional calculation within the DFT method was made (PSE
parametrization is described in details in [14]). In this work,
carbon nanotubes 7.95~{\AA} in diameter, 95.2~{\AA} in length, and
with chirality (10,0) were used. Each system was placed in a cubic
water box. The distance between the hybrid under study and walls
of the water box was not less than 12~{\AA}. Upon modeling,
periodic conditions were used. Upon modeling the pressure (1 bar),
the temperature (293~$^\circ$K) and the atom number in the system
were unchanged. At the modeling beginning, the energy of the
system was minimized during 1000 cycles. The VMD Program was
applied for the visualization of the results of calculations [15].

\begin{figure}
\includegraphics[width=5cm]{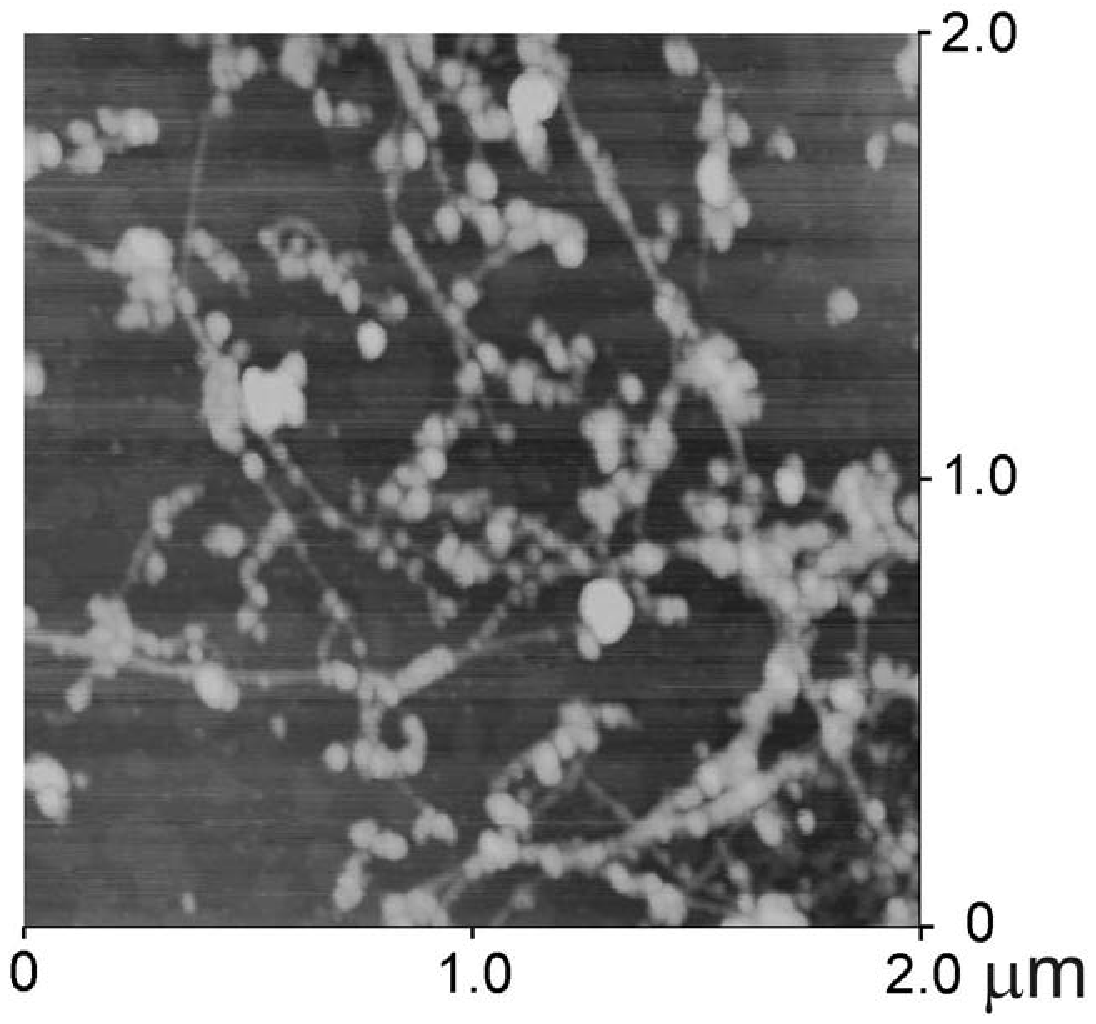}\\
{\large\it a}\\[2mm]
\includegraphics[width=5cm]{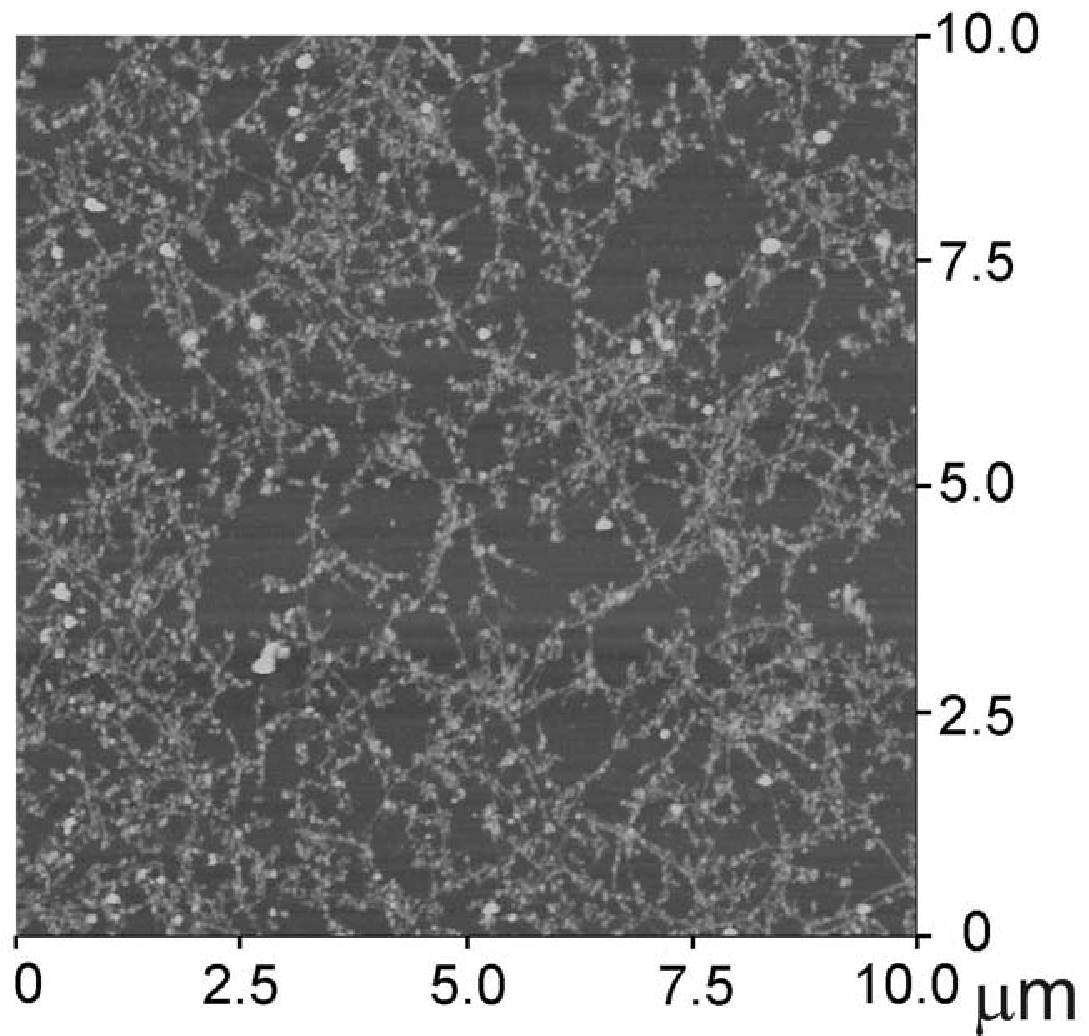}\\
{\large\it b} \vskip-3mm\caption{ AFM image of SWNT:PSE:GOX hybrids
on a mica substrate }
\end{figure}

\section{Results and Discussion }

\subsection{Analysis of the enzyme immobilization onto SWNTs by AFM}

We used the AFM method to study the morphology of the network of
SWNT deposited onto mica from a solution of nanotubes in
dichlorobenzene. Figure 1 shows the network of SWNTs obtained by the
spray method on mica. Both single nanotubes and their bundles can be
seen in Fig. 1. Then we deposited the molecular interface (PSE) on
the network, and finally the enzyme GOX was immobilized. The AFM
image of these hybrids is presented in Fig. 2. As is seen from Fig.
2, the GOX globular structures are mainly placed along nanotubes,
and their heights are in the 4.38--6.37-nm range. As the height of
one enzyme globule is about 4 nm [6], the obtained heights are the
sum of heights of a COX globule and a nanotube. The surface of some
nanotubes is fully covered with the enzyme. The detailed analysis of
hybrid heights revealed that they may be 8--8.5 nm in some cases,
and this indicates that the enzyme dimer is created on a nanotube.
Thus, it may be concluded that, due to the PSE molecular interface,
the enzyme is placed near the nanotube surface.

\begin{figure}
\includegraphics[width=7.5cm]{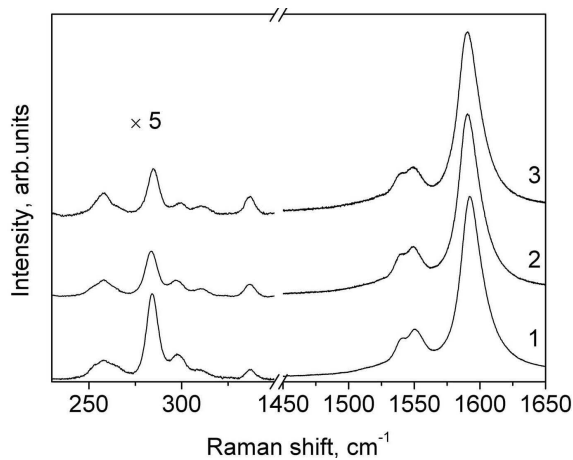}
\vskip-3mm\caption{ Raman spectra of SWNTs ({\it 1}), SWNT:PSE ({\it
2}) SWNT:PSE:GOX ({\it 3}) hybrids in the range of the RBM and G
modes. Each experimental spectrum  obtained by the laser excitation
at $\lambda _{\rm exc} = 632.8$ nm  }
\end{figure}

\subsection{Raman spectroscopy of carbon nanotubes in the PSE and GOX environments}

Information on the noncovalent interactions of organic molecules
with the nanotube surface can be obtained from the analysis of the
RR spectra of carbon nanotubes before and after the deposition of
these molecules [16]. The noncovalent interaction of a nanotube and
an organic molecule results in a shift of the nanotube vibrational
modes and in the intensity redistribution between bands. At first,
we studied such spectra of pristine nanotubes in bundles. Then we
measured the spectra of nanotubes with deposited PSE molecules and,
finally, after the immobilization of GOX molecules. Figure 3
presents the RR spectra of a nanotube network in the 225--1650
cm$^{-1}$ range before ({\it 1}) and after ({\it 2}) the deposition
of PSE, as well as after the addition of GOX ({\it 3}). It should be
noted that the spectra are similar, but the detailed analysis of
some bands reveals some slight change in the spectrum after the PSE
deposition onto the nanotube surface.

To analyze these changes in more details, we studied the most
informative fragments of the RR spectra of nanotubes: ranges of the RBM
(225--350 cm$^{-1})$ and G (1500--1650 cm$^{-1})$ modes.

\begin{figure}
\includegraphics[width=7.5cm]{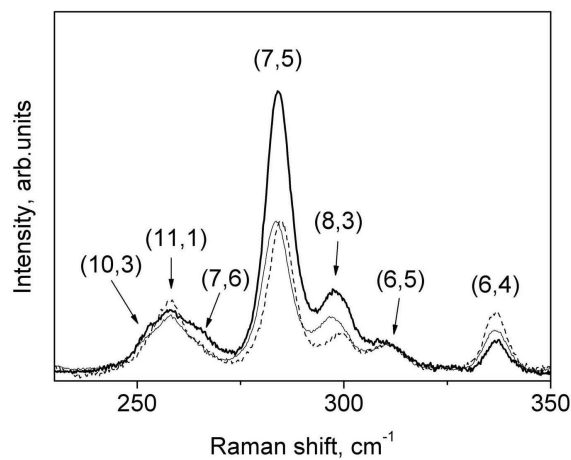}
\vskip-3mm\caption{ Low-frequency fragment of the spectra of SWNTs
(bold line), SWNT:PSE (thin line), and SWNT:PSE:GOX (broken line) in
the RBM range }
\end{figure}

\begin{table*}[!]
\noindent\caption{Position (\boldmath$\omega _{\rm RBM}$,
cm$^{-1}$), width at the half of the height ($\Delta \Gamma$,
cm$^{-1}$), integral intensity of Lorentzian curves normalized to
the band corresponding to the G$^{+ }$ band, chirality of nanotubes
(n,m), and second electronic transition of the semiconducting SWNTs
($E^{S}_{22}$, eV) located close to the excitation energy of a
He--Ne laser (1.96 eV)}\vskip3mm\tabcolsep10.0pt
\noindent{\footnotesize\begin{tabular}{c c c c c c c c c c c }
 \hline \multicolumn{1}{c}
{} & \multicolumn{1}{|c}{}& \multicolumn{3}{|c}{SWNT}&
\multicolumn{3}{|c}{SWNT:PSE}& \multicolumn{3}{|c}{SWNT:PSE:GOX}\\%
\cline{3-11}%
\multicolumn{1}{c}{(n,m)}& \multicolumn{1}{|c}{$E^{S}_{22}$, eV}&
\multicolumn{1}{|c}{}& \multicolumn{1}{|c}{$\Delta \Gamma $,}&
\multicolumn{1}{|c}{$S/S_{\rm G}^+$}& \multicolumn{1}{|c}{}&
\multicolumn{1}{|c}{$\Delta \Gamma $,}&
\multicolumn{1}{|c}{$S/S_{\rm G}^+$}&
\multicolumn{1}{|c}{$\omega$,}& \multicolumn{1}{|c}{$\Delta \Gamma
$,}&
\multicolumn{1}{|c}{$S/S_{\rm G}^+$}\\%
\multicolumn{1}{c}{}& \multicolumn{1}{|c}{[17]}&
\multicolumn{1}{|c}{$\omega$, cm$^{-1}$}&
\multicolumn{1}{|c}{cm$^{-1}$}&
\multicolumn{1}{|c}{$\times$100{\%}}& \multicolumn{1}{|c}{$\omega$,
cm$^{-1}$}& \multicolumn{1}{|c}{cm$^{-1}$}&
\multicolumn{1}{|c}{$\times$100{\%}}&
\multicolumn{1}{|c}{cm$^{-1}$}& \multicolumn{1}{|c}{cm$^{-1}$}&
\multicolumn{1}{|c}{$\times$100{\%}}\\%
\hline%
10.3 &1.90 &252.4 &7.4 &0.24 &251.4 &7.2 &0.25 &252.4 &5.8 &0.12  \\%
11.1 &2.00 &257.9 &8.6 &0.56 &257.9 &7.0 &0.40 &257.9 &7.4 &0.59  \\%
7.6 &1.90 &265.5 &7.4 &0.37 &264.1 &6.9 &0.22 &265.8 &7.1 &0.19  \\%
7.5 &1.88 &284.2 &7.1 &2.49 &283.6 &7.2 &1.36 &284.7 &6.7 &1.22  \\%
8.3 &1.87 &298.1 &9.6 &1.23 &297.3 &9.4 &0.59 &299.2 &8.8 &0.36  \\%
6.5 &2.18 &310.7 &7.2 &0.26 &310.8 &8.3 &0.21 &311.0 &6.9 &0.19  \\%
6.4 &2.09 &337.2 &4.9 &0.21 &336.7 &5.2 &0.29 &336.6 &4.7 &0.37  \\%
\hline
\end{tabular}}\vskip3mm
\end{table*}

\begin{figure*}
\includegraphics[width=7.5cm]{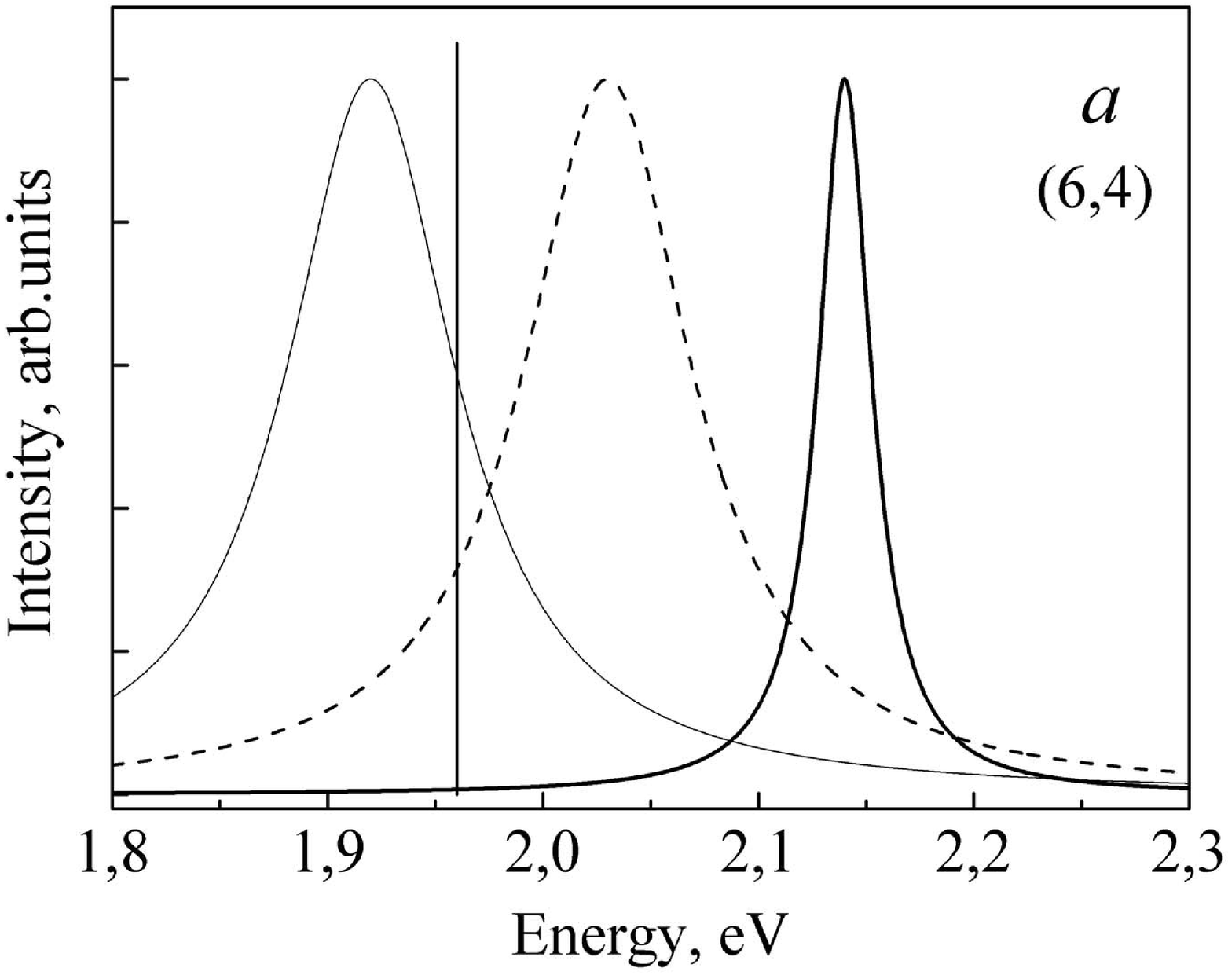}\hspace{1.5cm}\includegraphics[width=7.5cm]{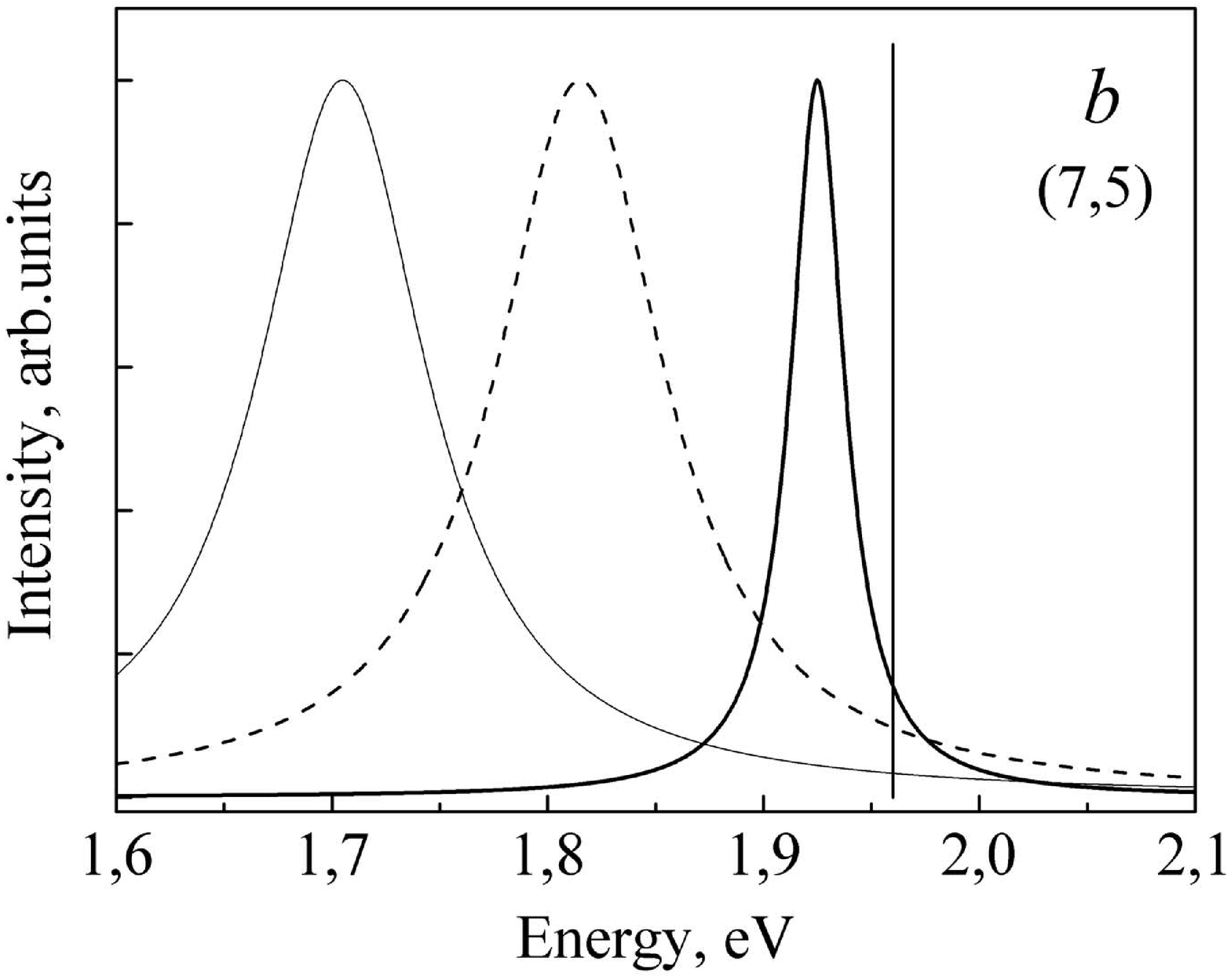}\\
{\large\it a\hspace{9cm}b} \vskip-3mm\caption{ Scheme of the
resonant windows of semiconducting nanotubes with different
chiralities (6,4) ({\it a}) and (7,5) ({\it b}) in the SDS
surrounding (solid curve), SWNTs in the film (thin line) and
SWNT:PSE (broken line). The energy (1.96 eV) of a He-Ne laser is
shown by the vertical line. Lorentzian curves were used to describe
the contours of resonant windows }
\end{figure*}

\subsubsection*{3.2.1 Radial breathing mode}

Figure 4 presents the low-frequency fragment of the RR spectrum of
nanotubes, in which the RBM is observed. At the excitation by a
He--Ne laser in the 225--350 cm$^{-1}$ range, seven intense bands
are observed for nanotubes obtained by the CoMoCat method. The
spectra were approximated with a sum of seven Lorentzians. The
frequency position of the peaks, area, and width at the half of the
height of bands for the both samples were determined. These results
are presented in Table~1. The nanotube chirality shown in Table 1
was determined earlier for these nanotubes  [17]. As follows from
Table 1, all bands were assigned to semiconducting nanotubes. The
spectra were normalized to the intensity of the high-frequency
component of the G-mode (G$^{+})$, which allowed us to compare the
band intensities in the spectra of three samples.

As follows from Table 1, the spectral shift in the RBM ($\sim $1.4
cm$^{-1})$ is observed for SWNT:PSE in comparison with SWNT bundles.
This shift is caused by the interaction of PSE molecules with
nanotubes. In addition, a decrease and an enhancement of the band
intensities are seen at 265.5, 284.2, 298.1 cm$^{-1}$ and at 257.9,
310.7, 337.2 cm$^{-1}$, respectively. It is known that the RBM bands
are very sensitive to resonance conditions, which are determined by
the location of the nanotube electronic level relative to the laser
energy. It is well known that the position of the nanotube
electronic level depends on the interaction with the environment
[18]. Thus, when the nanotube electronic level (in our case, it is
the second electronic level ($E^{S}_{22}))$ is located higher than
the energy of a laser, it becomes lower upon the environmental
interaction, the resonance conditions are improved, and the
intensity of the corresponding band increases. Otherwise, the
resonance conditions become worse, and the intensity of the
corresponding band decreases.

The resonance windows for two semiconducting nanotubes with
different positions of electronic levels relatively to the laser
energy (1.96 eV) are schematically shown in Fig. 5. The laser energy
is marked with the vertical line. Other lines are obtained using
Lorentz functions. In addition, the bands corresponding to the
aqueous suspension of nanotubes with a surfactant (sodium dodecyl
sulfate (SDS)) (thick line) and in the film (thin line) and their
hybrids with PSE (thin broken line) are shown in Fig. 5. The energy
of the electronic transition for nanotubes ($E^{S}_{22})$ in the
aqueous suspension of individual nanotubes with the SDS is by
60--160 meV higher,  than the energy of this level for nanotubes in
a film [18, 19]. This red shift is due to the strong interaction of
nanotubes in bundles, which are situated in a film. For SWNT:PSE,
the bundling effect is much smaller, because the organic molecules
become embedded between nanotubes in bundles in a solution during
the ultrasonication and, thus, decrease the interaction of
nanotubes. As a result, due to the interaction between nanotubes and
organic molecules, the $E^{S}_{22}$ level for such SWNTs is still
lower than that for individual nanotubes in an aqueous suspension.

\begin{table*}[!]
\noindent\caption{ Position (\boldmath$\omega _{\rm RBM}$,
cm$^{-1})$, width at the half of the height ($\Delta \Gamma$,
cm$^{-1})$, and integral intensities of curves described by
Lorentzian and BWF (with the asymmetry parameter $1/q$) functions
normalized to the total intensity of the band assigned to G band
($S/S_{\max})$}\vskip3mm\tabcolsep9.5pt
\noindent{\footnotesize\begin{tabular}{c c c c c c c c c c c c }
 \hline \multicolumn{4}{c}
{SWNT} & \multicolumn{4}{|c}{SWNT:PSE}&
\multicolumn{4}{|c}{SWNT:PSE:GOX}\\%
\hline%
\multicolumn{1}{c}{$\omega$,}& \multicolumn{1}{|c}{$\Delta\Gamma$,}&
\multicolumn{1}{|c}{$-1/q$}& \multicolumn{1}{|c}{$S/S_{\rm G}$}&
\multicolumn{1}{|c}{$\omega$,}&
\multicolumn{1}{|c}{$\Delta\Gamma$,}& \multicolumn{1}{|c}{$-1/q$}&
\multicolumn{1}{|c}{$S/S_{\rm G}$}& \multicolumn{1}{|c}{$\omega$,}&
\multicolumn{1}{|c}{$\Delta\Gamma$,}& \multicolumn{1}{|c}{$-1/q$}&
\multicolumn{1}{|c}{$S/S_{\rm G}$}\\%
\multicolumn{1}{c}{cm$^{-1}$}& \multicolumn{1}{|c}{cm$^{-1}$}&
\multicolumn{1}{|c}{}& \multicolumn{1}{|c}{$\times100{\%}$}&
\multicolumn{1}{|c}{cm$^{-1}$}& \multicolumn{1}{|c}{cm$^{-1}$}&
\multicolumn{1}{|c}{}& \multicolumn{1}{|c}{$\times100{\%}$}&
\multicolumn{1}{|c}{cm$^{-1}$}& \multicolumn{1}{|c}{cm$^{-1}$}&
\multicolumn{1}{|c}{}& \multicolumn{1}{|c}{$\times100{\%}$}\\%
\hline%
1521.7 &30.1 &0.17 &3.1 &1520 &29.2 &0.15 &3.9 &1518.4 &37.8 &0.17 &5.3  \\%
1539.7 &12.6 & &5.7 &1539.1 &12.2 & &6.1 &1539.4 &16.5 & &8.3  \\%
1551.2 &15.0 & &11.1 &1550.0 &14.6 & &10.3 &1550.5 &13.2 & &7.4  \\%
1592.2 &15.4 & &48.8 &1590.5 &15.1 & &46.5 &1590.0 &15.4 & &45.8  \\%
1601.4 &31.9 & &31.3 &1598.5 &31.4 & &33.2 &1598.0 &28.9 & &33.2  \\%
\hline
\end{tabular}}
\end{table*}

In our case, the nanotubes with (11,1), (6,5), and (6,4) chiralities have the electronic level
($E^{S}_{22})$, which is higher than
the laser energy. So, under the interaction of nanotubes with PSE, the
band intensity increases. This rule works unambiguously for nanotubes with
(6,4) chirality. Contrary
to this, the level $E^{S}_{22 }$ of nanotubes with (10,3), (7,6), (7,5), and
(8,3) chiralities is lower than the laser energy.
Therefore, under the interaction of these nanotubes with PSE, the band
intensity must decrease, which is observed experimentally.

We note that, after the adsorption of GOX molecules onto SWNT:PSE, the spectral
transformation is weaker. For example, the intensities of bands assigned to
(7,5), (10,3) and (7,6) nanotubes do not change practically.

\subsubsection*{3.2.2 Tangential mode}

The most intense band in the RR spectra of SWNTs lies in the
1550--1620~cm$^{-1}$ range and corresponds to the high-frequency
component of the tangential G mode [20]. This band for nanotubes and
their hybrids with PSE and PSE:GOX is shown in Fig. 6. As is seen
from Fig. 6, the intense band in the RR spectrum of the nanotubes in
bundles (1592.4~cm$^{-1})$ is low-shifted by 1.7~cm$^{-1 }$ in the
SWNT:PSE film. This shift takes place, because the interaction
energy of nanotubes in bundles decreases due to a weaker
nanotube-PSE binding energy. The intense band in the spectrum of the
SWNT:PSE:GOX complex has a peak at 1590.5~cm$^{-1}$, which indicates
the insignificant influence of GOX on the RR spectrum of
nanotubes:PSE (located at 1590.7~cm$^{-1})$.\looseness=1

\begin{figure}
\includegraphics[width=7.5cm]{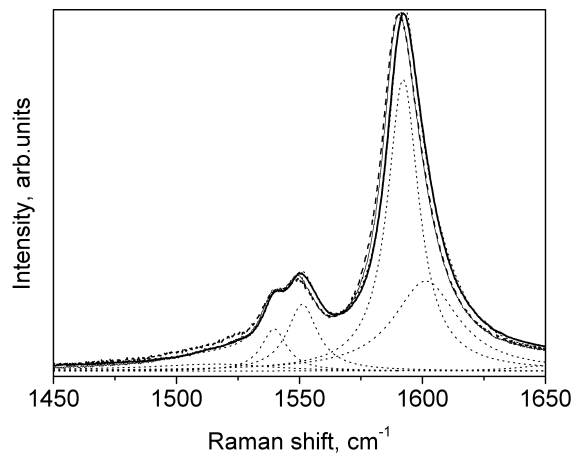}
\vskip-3mm\caption{ High-frequency fragment of SWNTs (bold line),
SWNT:PSE (thin line), and SWNT:PSE:GOX (broken line) hybrids in the
G mode range }
\end{figure}

Some differences in the band positions and intensities were obtained
after the fitting with the sum of approximation functions. Each
experimental spectrum has been fitted with a minimal number of
approximation functions (Fig.~6) consisting of the sum of four
Lorentzians and one Breit--Wigner--Fano (BWF) function, $I(\omega
)=I_{0}\{1 + (\omega -\omega _{0})/q\Gamma \}^{2}/\{1 + [(\omega
-\omega _{0})/ \Gamma ]^{2}\}$ [21], where $I_{0}$, $\omega _{0}$,
$\Gamma $, and $q$ are the intensity, BWF frequency, broadening
parameter, and asymmetry parameter, respectively. The BWF function
is used for describing the low-frequency spectral band, which
appears due to the presence of metallic nanotubes. On the basis of a
good coincidence between the experimental spectrum and the sum of
approximating curves, the parameters of these curves were determined
(Table~2). The total intensity of the whole spectral fragment was
set as 100{\%}. As is seen from Table~2, a low-frequency shift from
0.3 to 3.4~cm$^{-1}$ range was observed in the film with PSE for all
bands of nanotubes.

Thus, in accordance with changes in the RR spectra of nanotubes in the PSE
environment relative to the spectrum of pristine nanotubes, we can
conclude that the PSE molecule forms a noncovalent hybrid with a carbon
nanotube, which favors the further immobilization of enzyme molecules. At the same
time, the absence of significant changes in the RR spectrum of the SWNT:PSE:GOX
hybrid (as compared with the SWNT:PSE one) evidences that these PSE molecules
isolate efficiently the GOX molecules from nanotubes, preventing their
direct interaction with the nanotube surface.

\subsection{Structure and interaction energy of the SWNT:PSE:GOX complex
in the water environment: molecular dynamics modeling}

While modeling the SWNT:PSE:GOX complex, we were aimed to reveal
whether one PSE molecule is able to keep one GOX molecule
(covalently bound to it) near the nanotube surface. The other task
was to determine the interaction energies between the PSE:GOX
complex and SWNTs. It is known that PSE reacts with amino groups
of peptides [8[. In this case, the CO-NH bond is formed. One PSE
molecule was attached to one of the lysine side residues of GOX.
As a result, a new compound was obtained (denoted as PSE:GOX). It
consists of the pyrene fragment and a GOX molecule joined with a
linker. The PSE:GOX complex was connected to the SWNT surface by
the pyrene fragment at a distance of 3.5~{\AA}. The system was
minimized during 1000 cycles and then equilibrated for 10~ns with
a 1-fs step. After modeling the total energy of interaction
between SWNT and the whole PSE:GOX complex, the interaction
energies of SWNT and various parts of this complex were
determined. They are shown in Fig.~7. The equilibrated structure
of the SWNT:PSE:GOX complex is presented in Fig.~8.

\begin{figure}
\includegraphics[width=7.5cm]{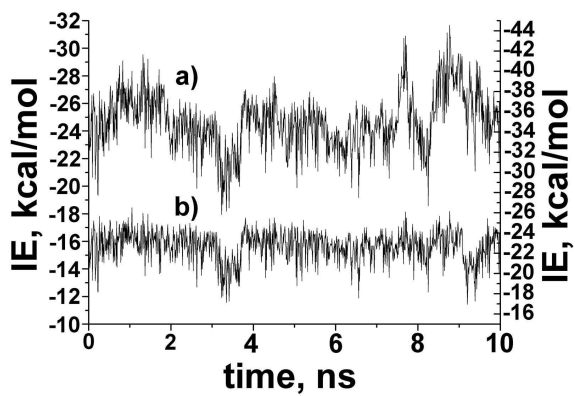}
\vskip-3mm\caption{ Interaction energies between PSE and SWNT ({\it
a}, left scale) and PSE-GOX and SWNT ({\it b}, right scale) }
\end{figure}

As is seen from Fig. 7, the energy of the interaction between the
pyrene fragment and the nanotube surface in the SWNT:PSE:GOX complex
(curve {\it a}) is in the interval ($-20$\,--\,$-25$) kcal/mol. This
result agrees very well with the interaction energy between pyrene
and nanotubes calculated for the SWNT:PSE complex [14]. Moreover,
Fig. 7 demonstrates that, during the whole period of modeling, the
pyrene fragment is strongly attached to the nanotube. This evidences
that one pyrene ``anchor'' is able to hold efficiently the GOX
enzyme on the nanotube. The total interaction energy between a
nanotube and the PSE:GOX complex (Fig. 7, curve {\it b}) (from $-22$
to $-27$ kcal/mol) is equal to the sum of energies of interactions
between PSE and a nanotube, as well as between GOX and a nanotube.
Thus, we can determine the energy of interaction between GOX and a
nanotube, by subtracting the SWNT:PSE interaction energy from the
total interaction energy (Fig. 7, curve {\it b}). Thus, the energy
of interaction between GOX and a nanotube changes during the
modeling from $-5$ to $-2$ kcal/mol. This energy is due to the
interactions between a nanotube and outward side residues of a GOX
molecule. The mutual orientation of GOX and the nanotube surface
changes slowly during the modeling. As a result, the GOX--nanotube
interaction energy changes as weakly. It may be concluded that the
modeling of the SWNT:PSE:GOX complex demonstrates the absence of
strong interactions between GOX and the nanotube surface. This fact
allows us to suppose that the GOX activity is not changed under the
formation of the SWNT:PSE:GOX complex.

\begin{figure}
\includegraphics[width=7.5cm]{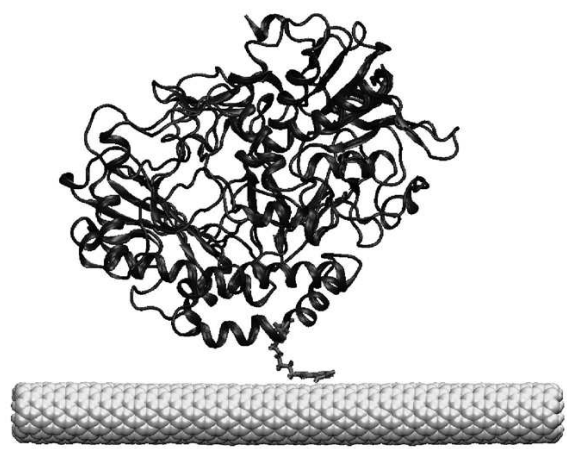}
\vskip-3mm\caption{ Equilibrated structure of the SWNT-PSE-GOX
complex. Water molecules are not shown }
\end{figure}

\begin{figure}
\includegraphics[width=7.5cm]{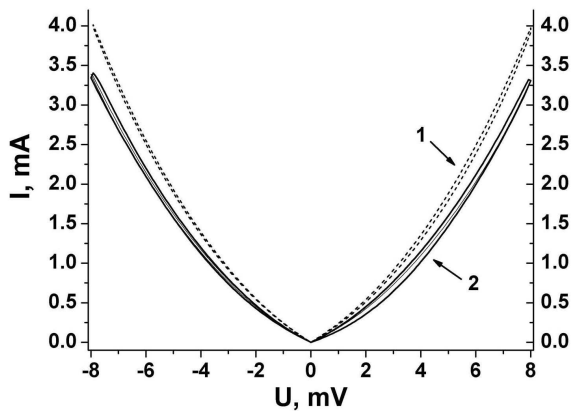}
\vskip-3mm\caption{ Current versus voltage for an SWNT network
placed between two gold contacts separated by 10 $\mu $m gap (curve
{\it 1}). Curve {\it 2} corresponds to the conductivity of a sample
after the adsorption of PSE  }
\end{figure}

\subsection{Conductivity of single-walled carbon nanotube
networks: effects of environment}

We carried out studies of the conductive properties of the
single-walled carbon nanotube network sprayed onto a quartz
substrate from their solution in dichlorobenzene. The dependence of
the current through the carbon nanotube network in the dry state at
variations of the voltage $(U)$ between two contacts is shown in
Fig. 9 (dotted line). The voltage is changed in (--8 -- +8) V range,
the maximum current $(I)$ runs up to 4 mA, and the $I(U)$ dependence
has nonlinear character. Most likely, the nonlinearity is related to
Schottky barriers, which originate at the contact between nanotubes
and a gold contact or between nanotubes of different conductivities
[22]. In addition, the $I$($U$) dependences manifest a small
hysteresis. To avoid the effects of a solution on the volt-ampere
characteristics, the presented dependences were obtained in 4--6
days after the fabrication of a nanotube network or after the
deposition of biomolecules.\looseness=1

In this study, the effects of bioorganic compounds (PSE and GOX)
deposited on the carbon nanotube network on its conductivity have
been investigated. After the deposition of the molecular interface
(PSE) from methanol and the drying of the film, its conductivity was
about 20{\%} less than the initial value (Fig. 9, solid line). Most
probably, such a decrease is caused by adsorbed PSE molecules, which
induce the appearance of scattering centers for charge carriers on
the nanotube surface, as it was observed earlier [2]. It should be
noted that the following GOX adsorption (Fig. 9, inside curve {\it
2}) has practically no effect on the conductivity of the nanotube
network. It is obvious that this molecular interface (PSE) isolates
the enzyme from the nanotube surface rather efficiently. The
appearance of the hysteresis for the $I$($U$) dependence in similar
measurements was explained earlier by the presence of charge traps
on the SiO$_{2}$ surface. These traps are occupied under the passage
of a current in one direction and are not depleted yet, when the
current has the opposite direction [19].

\begin{figure}
\includegraphics[width=7.5cm]{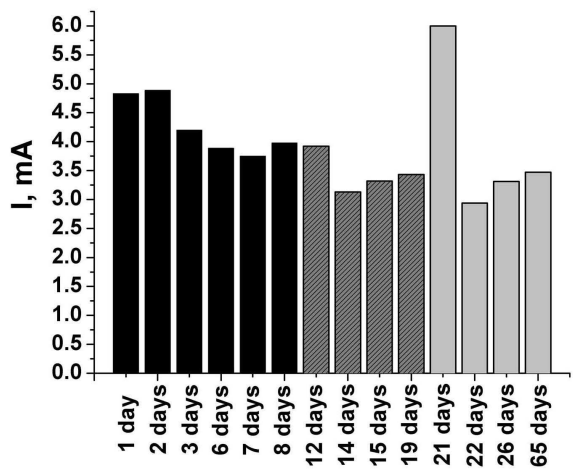}
\vskip-3mm\caption{ Time dependence of the maximal current (obtained
at 8~V) for a SWNT network placed between two gold contacts
separated by 10 $\mu $m gap (black column). Dashed and shaded
columns correspond to the current through the network measured after
the deposition of PSE and the GOX immobilization, respectively }
\end{figure}

In addition, the time dependence was built for the current passing through the
nanotube network, beginning from the nanotube network fabrication up
to the PSE deposition and the immobilization of GOX (Fig. 10). As is seen from
Fig. 10, some spread of current values is observed right away after the
nanotube network treatment with a bioorganic substance, which is
caused by drying the network after its wetting with methanol or water. After
some days, this spread of values becomes narrower. A change in the conductivity
is especially noticeable after the treatment of the nanotube network with water. At
once after the water evaporation, the film conductivity rises appreciably. Then
it reduces in the course of the time, but this takes a rather long time (some days at room
temperature).

\section{Conclusion }

The efficient immobilization of GOX onto a carbon nanotube network through the
molecular interface formed by PSE is carried out. This conclusion is based on
the analysis of AFM images of the network with the adsorbed enzyme, whose
globules locate mainly along a nanotube.

The band corresponding to the high-frequency component of the G mode in the RR
spectrum of the nanotube with adsorbed PSE is downshifted by 0.7 cm$^{-1}$
relative to this band in the spectrum of pristine nanotubes. The analysis of
the intensities of bands assigned to the RBM of nanotubes with adsorbed PSE in
comparison with the spectrum of pristine SWNTs revealed the intensity
transformation, which can be explained by a change of the resonance
condition with variation of the laser energy. Thus, we concluded that PSE molecules create
nanohybrids with SWNTs, which ensures the further enzyme immobilization. As the RR
spectrum of an SWNT:PSE:GOX film does not essentially differ from SWNT:PSE
ones, this indicates that the molecular interface (PSE) isolates the enzyme from
nanotubes strongly enough.

Our studies on the conductive properties of a single-walled carbon
nanotube network sprayed onto a quartz substrate from a solution of
nanotubes in dichlorobenzene demonstrated that the $I$($U$)
dependence has nonlinear character. Most likely, the nonlinearity is
related to Schottky barriers, which originate on the contact between
nanotubes and the gold electrode, as well as between nanotubes with
different conductivities.

The deposition of bioorganic compounds (PSE and GOX) on the carbon
nanotube network is accompanied by a decrease of their conductivity.
Most probably, such a decrease is caused by adsorbed PSE molecules,
which induce the appearance of scattering centers for charge
carriers on the nanotube surface. The following GOX adsorption has
practically no effect on the conductivity of the nanotube network
that evidences the reliable isolation of the nanotube surface from
the enzyme by means of the molecular interface (PSE). While studying
the properties of carbon nanotube networks, whose surface underwent
the treatment with a solution, it is necessary to consider the
gradual desorption of a solvent from the nanotube surface
(especially, this concerns water).\looseness=-1

\vskip3mm The authors acknowledge the Computational Center at B.
Verkin Institute for Low Temperature Physics and Engineering of
the National Academy of Sciences of Ukraine.

\vskip-3mm

\rezume{%
IММОБІЛІЗАЦІЯ ГЛЮКОЗООКСИДАЗИ НА СІТКУ\\ ОДНОСТІННИХ ВУГЛЕЦЕВИХ
НАНОТРУБОК}{В.О. Карачевцев, О.Ю. Гламазда, Є.С. Заруднєв,\\ М.В.
Карачевцев, В.С. Леонтьєв, О.С. Лінник,\\ О.С. Литвин, О.М.
Плохотниченко, С.Г. Степаньян} {При створенні біологічних сенсорів з
використанням одностінних вуглецевих нанотрубок (ОВНТ) треба
вирішити таку важливу проблему, як іммобілізація молекули, яка
повинна розпізнати мішень, на поверхні нанотрубок. В даній роботі
проведена іммобілізація ферменту глюкозооксидаза (ГОК) на поверхню
сітки нанотрубок, яка була одержана шляхом осадження нанотрубок з їх
розчину у діхлорбензолі за допомогою спрей-методу. У ролі
молекулярного інтерфейсу було застосовано сукцинімідний ефір
1-піренбутанової кислоти (ПСЕ), біфункціональна молекула якого
забезпечує хімічний зв'язок з оболонкою ферменту, а друга її частина
(піренова) адсорбується на поверхню нанотрубки. Використання такого
молекулярного інтерфейсу виключає, з одного боку, пряму адсорбцію
ферменту на поверхню нанотрубки, яка знижує його активність, а з
другого, забезпечує локалізацію ферменту поблизу нанотрубки.
Порівняння спектрів резонансного комбінаційного розсіювання світла
(РКРС) нанотрубок з їх спектром в оточенні ПСЕ вказує на створення
наногібриду молекулою ПСЕ з нанотрубкою, що дає підставу для
подальшої іммобілізації ферментів. Оскільки спектри РКРС плівок
ОВНТ:ПСЕ:ГОК суттєво не відрізняються від спектрів ОВНТ:ПСЕ, то
можна стверджувати, що молекулярний інтерфейс ПСЕ достатньо міцно
ізолює фермент від нанотрубки. Ефективна іммобілізація ферменту ГОК
поблизу вуглецевої нанотрубки завдяки ПСЕ підтверджується за
допомогою зображень, отриманих атом-силовим мікроскопом. Молекулярна
динаміка дозволила встановити структури отриманих нанобіогібридів та
енергії міжмолекулярної взаємодії між компонентами потрійного
комплексу у водному оточенні. Було також досліджено провідні
властивості сітки ОВНТ з адсорбованими молекулами ПСЕ та ГОК.
Адсорбція молекул ПСЕ на сітку з ОВНТ супроводжується зменшенням
провідності, яке, скоріш за все, пов'язано\linebreak з появою
розсіювальних центрів для носіїв заряду у нанотрубках.}

\end{document}